\documentclass[twocolumn]{aastex631}
\shorttitle{Glitch clustering phenomenon}
\shortauthors{Zhu et al.}
\graphicspath{{./}{figures/}}
\usepackage{color}

\begin{document}

\title{Glitches and glitching clusters in rotation-powered pulsars}

\correspondingauthor{Xiao-Ping Zheng}
\email{zhxp@ccnu.edu.cn}

\author[0009-0002-0748-2105]{Pei-Xin Zhu}
\affiliation{School of Physics, Huazhong University of Science and Technology, Wuhan 430074, People's Republic of China}

\author[0000-0001-8868-4619]{Xiao-Ping Zheng}
\affiliation{Institute of Astrophysics, Central China Normal University, Wuhan 430079, People's Republic of China}
\affiliation{ Department of Astronomy, Huazhong University of Science and Technology, Wuhan 430074, People's Republic of China}



\begin{abstract}
The study of pulsar glitch phenomena serves as a valuable probe into the dynamic properties of matter under extreme high-density conditions, offering insights into the physics within neutron stars. Providing theoretical explanations for the diverse manifestations observed in different pulsars has proven to be a formidable challenge. By analyzing the distribution of glitch sizes and waiting times, along with the evolution of cumulative glitch sizes over time, we have uncovered a long-term clustering phenomenon for pulsar glitches. This perspective allows us to approach the distinct glitch representations in various pulsars from a unified standpoint, connecting the same periodicity of observational data to the randomness. Without relying on specific physical models, we utilized the coefficient of variation to numerically determine optimal clustering numbers and clustering periods for sample pulsars. Our analysis involving 27 pulsars has revealed a clear linear relationship between the glitch cluster period and characteristic age. Of interest, the cumulative distribution of functions of cluster sizes and interval times have the same patterns,  which can be synchronously fitted by Gaussian processes.  These results may indicate novel understandings of glitches and the resulting processes.

\end{abstract}



\section{Introduction} \label{sec:intro}
Pulsars are commonly recognized as highly magnetized rotating neutron stars, with their rotation frequencies steadily decreasing due to various mechanisms such as magnetic dipole radiation \citep{gunn1969magnetic,pacini1968rotating}. These celestial objects exhibit remarkable precision in their rotation rates, rivaling that of atomic clocks, especially in the case of millisecond pulsars \citep{ryba1991high,cognard1995high}. However, under certain exceptional circumstances, their rotation frequency can experience a sudden increase $\Delta \nu$, often accompanied by changes spin-down rate $\Delta \dot\nu$. These spin-up events are referred to as ``glitches" \citep{radhakrishnan1969detection,reichley1969observed}. Following a rapid increase in rotation frequency over a short period, there typically ensues an approximately exponential recovery process, spanning from a few days to several hundred days, during which the pulsar returns to its state before the glitch occurrence \citep{yu2013detection}. Benefiting from over 50 yr of continuous pulsar observations, the number of observed glitches has steadily increased, with 671 glitches detected in a sample of 224 pulsars. Glitches exhibit a wide range of parameters, with $\Delta \nu/\nu \sim 10^{-11}-10^{-5}$ and $\Delta \dot{\nu}/\dot{\nu} \sim 10^{-5}-10^{-2}$, while the time intervals between adjacent glitches fall within the range of 20-$10^{4}$ days \citep{eya2019distributions}.

The prevailing theoretical framework suggests that glitches result from angular momentum transfer between the faster-rotating superfluid component in the interior and the solid outer crust of a neutron star \citep{anderson1975pulsar,alpar1977pinning,alpar1984vortexb,alpar1984vortexa,alpar1989vortex}. Due to the action of electromagnetic braking torques, the rotational velocity of the solid crust gradually decreases, while the superfluid vortices maintain a higher rotational speed. The velocity difference between these two components stores angular momentum. Under appropriate conditions, superfluid vortices unpin and then repin, transferring angular momentum to the outer crust, leading to a glitch event.

The observations of the Vela pulsar (PSR J0835$ - $4510) are widely considered to align closely with the standard scenario. This is attributed partly to the successful explanation of postglitch relaxation observed by \citet{alpar1984vortexb,alpar1984vortexa}, and partly to the achievement of angular momentum transfer from the superfluid component to the pulsar's crust. This transfer was initially evidenced by \citet{link1999pulsar}, who found a linear relationship between cumulative angular momentum and time span using glitch data spanning over 30 yr from the Vela pulsar. Quantitative calculations can be performed using the following formula:
\begin{displaymath}
\frac{I_{\mathrm{res}}}{I_c} \geq \frac{\bar{\Omega}}{|\dot{\Omega}|} A \equiv G,
\end{displaymath}
where $ I_{\mathrm{res}} $ is the moment of inertia of the angular momentum
reservoir,  ${I_c}$ the moment of inertia of the solid crust plus
any portions of the neutron star tightly coupled to it, and $ \bar{\Omega} $ and $ \dot{\Omega}$
denote the average spin rate and spin-down rate of the crust, respectively.
The slope $ A $ represents the ratio of cumulative relative glitch size to the observation time. Given that the observed values of coupling parameter $ G $ are consistent with theoretical calculations of the neutron star's crust \citep{ravenhall1994neutron}, it is commonly inferred that the superfluid reservoir resides within the inner crust. Building on the concept of the angular momentum reservoir, several studies have explored the calculation of the ratio of crustal fluid moment of inertia to total moment of inertia to constrain the mass and radius of the neutron star \citep{delsate2016giant,montoli2020role}.

The analysis conducted for the Vela pulsar \citep{lyne1996very} was statistically extended to all pulsars that have been monitored over many years for glitch activity, including those that have not glitched \citep{lyne2000statistical,fuentes2017glitch}. \citet{lyne2000statistical} collected data from 279 pulsars with spin frequency derivative, $\dot{\nu}$, ranging from $ 10^{-16}$ to $10^{-9} \text{ s}^{-2}$, encompassing 49 glitches from 18 pulsars for statistical analysis. They defined the glitch spin-up rate, $\dot{\nu}_g$, as
\begin{displaymath}
	\dot{\nu}_{\rm{g}} = \frac{\sum_i\sum_j \Delta \nu_{ij}}{\sum_i T_i} ,
\end{displaymath}
where the frequency change $ \Delta \nu_{ij} $ indicates the glitch $ j $ of the pulsar $ i $, and $ T_i $ is the interval time between consecutive events of pulsar $ i $. Comparisons were made between the cumulative acceleration of the crust due to the size of each glitch over a long time and the spin-down over the same duration to obtain a linear increase in glitch activity with the increasing spin-down rate of pulsars. This linear relationship revealed, except for the youngest
pulsars (including the Crab), with $\tau_{\mathrm{c}}<10^4 \mathrm{yr}$, a ratio of 0.017 between a pulsar's glitch spin-up rate and its spin-down rate. This implies that 1.7$\% $ of the moment of inertia of a typical neutron star is involved in releasing an excess angular moment at a glitch, as evidenced by studies of the Vela pulsar. Further studies expanding the sample confirmed these findings but reported slightly different ratios \citep{espinoza2011study,fuentes2017glitch}. Recently, \citet{basu2022jodrell} added 106 new glitches in 70 radio pulsars into the sample to present  a mean reversal of 1.8$\% $ of the spin-down as a consequence of glitches. These findings, both in multiple pulsars and individual objects, statistically demonstrate consistency with the required unpinning superfluid of pulsar glitches.

\begin{deluxetable*}{llccccccccccc}
	\tablecaption{Basic Information for 32 pulsars with no fewer than five glitches.\label{tab:information}}
	\tablehead{
		\colhead{J$ - $name} & \colhead{B$ - $name} & \colhead{N} & \colhead{$A$} & \colhead{$G$} & \colhead{$\nu$} & \colhead{$\dot{\nu}$} & \colhead{$\tau_{c}$} & \colhead{$ \Delta $T\_seq} & \colhead{RS\_seq} & \colhead{$ \Delta $T\_seq} & \colhead{RS\_seq} & $ \Delta S $ \\ 
		~ & ~ & ~ & $10^{-9}$ d${}^{-1}$ & \% & Hz & $10^{-12}$ Hz s${}^{-1}$ & kyr & normal & normal & possion & possion & $ \% $}
	\startdata
	J1341$ - $6220 & B1338$ - $62 & 33 & 2.09  & 1.85  & 5.17  & 6.77  & 12.14  & 5 & 1 & 1 & 1 & 14.47   \\ 
	J0537$ - $6910 & ~ & 53 & 2.48  & 0.89  & 62.03  & 199.23  & 4.95  & 10 & $ * $ & 1 & 1 & 0.80   \\ 
	J0835$ - $4510 & B0833$ - $45 & 24 & 1.97  & 1.63  & 11.19  & 15.67  & 11.35  & 10 & 10 & 1 & 1 & 0.67   \\ 
	J1420$ - $6048 & ~ & 7 & 1.54  & 1.47  & 14.67  & 17.89  & 13.02  & $ * $ & $ * $ & 1 & 1 & 0.54   \\
	J1740$ - $3015 & B1737$ - $30 & 37 & 0.81  & 1.22  & 1.65  & 1.27  & 20.69  & 1 & 1 & $ * $ & 1 & 8.63   \\
	J1413$ - $6141 & ~ & 14 & 1.14  & 1.14  & 3.50  & 4.09  & 13.61  & 1 & 1 & 5 & 15 & 5.59   \\ 
	J0205$ + $6449 & ~ & 9 & 1.71  & 0.67  & 15.22  & 44.87  & 5.39  & 15 & 5 & $ * $ & 10 & 6.29   \\ 
	J1048$ - $5832 &   B1046$ - $58 & 9 & 1.03  & 1.54  & 8.08  & 6.28  & 20.45  & $ * $ & 2.5 & $ * $ & 1 & 6.17   \\
	J1826$ - $1334 &  B1823$ - $13 & 7 & 0.90  & 1.41  & 9.85  & 7.31  & 21.43  & $ * $ & 10 & $ * $ & 1 & 7.98   \\ 
	J1841$ - $0524 & ~ & 7 & 0.44  & 0.97  & 2.24  & 1.18  & 30.30  & $ * $ & $ * $ & $ * $ & 10 & 5.69   \\ 
	J1709$ - $4429 & B1706$ - $44 & 5 & 0.82  & 1.05  & 9.76  & 8.86  & 17.51  & $ * $ & $ * $ & $ * $ & $ * $ & 8.25 \\
	J1357$ - $6429 & ~ & 5 & 1.76  & 0.94  & 6.02  & 13.05  & 7.33  & $ * $ & $ * $ & $ * $ & 10 & 6.25   \\ 
	J1801$ - $2304 & B1758$ - $23 & 15 & 0.22  & 0.96  & 2.40  & 0.65  & 58.50  & $ * $ & 2.5 & 15 & 1 & 1.29   \\ 
	J2229$ + $6114 & ~ & 9 & 0.74  & 0.56  & 19.37  & 29.37  & 10.48  & $ * $ & 15 & 10 & $ * $ & 3.96   \\
	J1023$ - $5746 & ~ & 7 & 3.92  & 1.32  & 8.97  & 30.88  & 4.62  & $ * $ & $ * $ & 2.5 & 5 & 2.01   \\ 
	J1801$ - $2451 &   B1757$ - $24 & 7 & 1.40  & 1.59  & 8.00  & 8.20  & 15.52  & $ * $ & $ * $ & 2.5 & 15 & 1.14   \\ 
	J1803$ - $2137 &  B1800$ - $21 & 6 & 1.70  & 1.96  & 7.48  & 7.52  & 15.81  & $ * $ & $ * $ & 15 & 2.5 & 1.97  \\
	J0631$ + $1036 & ~ & 17 & 0.57  & 1.81  & 3.47  & 1.26  & 43.68  & 1 & 1 & $ * $ & 1 & 18.60   \\ 
	J0742$ - $2822 & ~ & 9 & 0.26  & 2.95  & 6.00  & 0.60  & 157.51  & $ * $ & 1 & $ * $ & 1 & 11.89   \\ 
	J1731$ - $4744 &   B1727$ - $47 & 6 & 0.26  & 1.53  & 1.21  & 0.24  & 80.57  & $ * $ & 1 & 15 & 1 & 10.90   \\ J1952$ + $3252 & ~ & 6 & 0.40  & 3.12  & 25.30  & 3.74  & 107.45  & $ * $ & 1 & 15 & 1 & 11.08   \\ 
	J0729$ - $1448 & ~ & 6 & 1.17  & 3.01  & 3.97  & 1.79  & 35.29  & $ * $ & 1 & 2.5 & 1 & 54.31   \\ J1105$ - $6107 & ~ & 5 & 0.31  & 1.44  & 15.82  & 3.97  & 63.37  & $ * $ & $ * $ & $ * $ & 2.5 & 4.63   \\ 
	J1617$ - $5055 & ~ & 5 & 0.12  & 0.07  & 14.42  & 28.09  & 8.16  & 5 & 10 & 2.5 & 10 & 6.30   \\ J1708$ - $4008 &  & 6 & 3.05  & 1.99  & 0.09  & 0.16  & 8.92  & $ * $ & 10 & $ * $ & $ * $ & 10.85   \\ 
	J1737$ - $3137 & ~ & 7 & 0.84  & 3.16  & 2.22  & 0.68  & 51.57  & $ * $ & 1 & 10 & 1 & 13.37   \\ J1825$ - $0935 & ~ & 7 & 0.02  & 0.35  & 1.30  & 0.09  & 233.36  & 10 & 1 & $ * $ & 1 & 0.62   \\ 
	J0534$ + $2200 & B0531$ + $21 & 30 & 0.07  & 0.01  & 29.95  & 377.54  & 1.26  & 1 & 1 & $ * $ & 1 & 3.63   \\ 
	J1814$ - $1744 & ~ & 7 & 0.02  & 0.12  & 0.25  & 0.05  & 84.82  & $ * $ & $ * $ & 15 & $ * $ & 0.13   \\ 
	J2021$ + $3651 & ~ & 5 & 1.30  & 1.64  & 9.64  & 8.89  & 17.22  & $ * $ & $ * $ & 10 & $ * $ & 1.87   \\ 
	J1902$ + $0615 & ~ & 6 & $ \sim 10^{-4} $ & 0.02  & 1.48  & 0.02  & 1387.80  & 10 & $ * $ & 15 & 2.5 & $ \sim 10^{-3} $  \\ 
	J2225$ + $6535 &  B2224$ + $65 & 5 & $ \sim 10^{-5} $ & $ \sim 10^{-3} $ & 1.47  & 0.02  & 1122.48  & $ * $ & 2.5 & $ * $ & 1 &   $ \sim 10^{-3} $\\ 
	\enddata
	\tablecomments{N represents the number of glitches occurring, $A$ denotes the slope fitted to the evolution of cumulative glitch size over time, $G$ represents the coupling constant, and $\tau_{c}$ stands for the characteristic age of the pulsar. $ \Delta $T\_seq and RS\_seq represent the sequence of time intervals between adjacent glitches and the sequence of relative sizes of the glitches, respectively, for this pulsar. We conducted an Anderson-Darling (AD) test on the intervals between individual glitches and their relative sizes, indicating the minimum significance levels for rejecting normal or homogeneous Poisson distributions. Instances marked with an asterisk $ * $ suggest that we cannot reject the hypotheses of Gaussian or homogeneous Poisson distributions. The final column presents the normalized dimensionless area error parameter for each pulsar.}
\end{deluxetable*}

The superfluid model is expected to encompass other reservoir effects, specifically referring to some glitching regularities resulting from the superfluid reservoir's role in storing and transferring angular momentum, such as glitches recurring quasiperiodically, exhibiting a narrowed spread in size, and displaying a size-waiting-time correlation. In Vela and PSR J0537$ - $6910, the glitches recur quasiperiodically. PSR J0537$ - $6910 is the only case where a correlation between glitch size and waiting time to the next glitch exists \citep{middleditch2006predicting}. However, this manifestation does not exist in the rest of the population, suggesting that there is no reservoir effect. To address this challenge, numerous studies have focused on analyzing the distributions of glitch sizes and waiting times, as well as their relationship \citep{melatos2008avalanche,fulgenzi2017radio,howitt2018nonparametric,carlin2020long}, and have attempted to resolve this issue by modeling various triggering mechanisms \citep{ruderman1969neutron,andersson2003pulsar,cheng1988spontaneous}. \cite{melatos2008avalanche} demonstrated that the size distributions follow a power law distribution, albeit with variations in the index across different pulsars, while the waiting-time distributions adhere to an exponential pattern. \cite{fuentes2019glitch} reanalyzed data from eight pulsars with at least ten detected events, finding that, except for Vela and PSR J0537$ - $6910, the size distributions in the remaining six pulsars are best described by power-laws, exponentials, or log-normal functions, with waiting-time distributions best fitted by exponentials. Two models, star quakes model indicating that the crust fractures and readjusts its structure at the critical strain and snowplow model implying a glitch occurs when the critical lag reaches the maximum pinning force, predict a correlation between glitch size and waiting time, whereas others, such as self-organized critical systems, explain systems where the size distribution follows a power law and the waiting-time distribution follows an exponential \citep{jensen1990lattice}. \cite{fuentes2017glitch} argued the observed bimodal distribution of glitch sizes to distinguish between large and small glitches, with the boundary at $\Delta \nu =10\mu$ Hz. Furthermore, \cite{fuentes2019glitch} suggested that both large and small glitches draw from the same angular momentum reservoir but may be  triggered by difference mechanisms, with large glitches occurring once a critical state is reached, while small glitches arise more randomly.

This work is intended to concentrate solving the periodicity of glitches on long timescales and the equivalence of a unified mechanism underlying the diverse glitch manifestations in different pulsars within glitch ensembles, while maintaining the integrity of the glitch data itself. We initiated our investigation by examining the size distribution of glitches and the changes in cumulative relative size over time. This led to a discovery of temporal clustering phenomenon in pulsar glitches; as a matter of fact, previous work had shown the signs of clustering in the Crab Pulsar by testing the deviations from a homogeneous Poisson process \citep{carlin2019temporal}. We hence consider glitch clusters instead of a single glitch a unified perspective to understand the diverse representations in different pulsars, shedding light on the relationship between the several periodic sources and the other randomness in the distribution of glitch sizes and waiting times. Throughout this study, we refrain from relying on specific physical models and focus solely on exploring the glitch clustering  from a data analysis perspective. Glitch clusters are more liable to emerge periodically than repeated single glitches.  This will help to understand the
periodic accumulation of angular momentum with a random release.

The remainder of this paper is organized as follows. In Section \ref{sample}, we present the basic information for a sample of 32 pulsars, each exhibiting at least five glitches, and offer an overview of the sizes of glitches and their corresponding waiting times for various pulsars.
Section \ref{clusters} focuses on detailing the discovery of the clustering phenomenon of glitches over extended timescales and the numerical methods employed to determine long-term periodicity. 
In section \ref{statistical}, we statistically analyze data of glitch clusters, presenting cumulative distribution functions (CDFs) of cluster sizes and interval times between consecutive clusters, and we present a prominent linear relationship between the glitch cluster period and the characteristic age for 27 pulsars. 
In Section \ref{Conclusion}, we discuss the classification of the four types of pulsars introduced at the outset of the article. We begin by examining the differences in periodicity and randomness exhibited by glitches across various pulsars, which serve as the foundation for understanding the concept of glitch clustering. Additionally, we offer a rough estimate of $ G $ based on the assumption of partial release, and we discuss new perspectives on the superfluid reservoir and the observed $ G $ within the context of the timescale associated with glitch clustering.
Finally, we provide a summary of the study and discuss the possible implications for physical understandings.

\begin{figure*}
	\centering
	\includegraphics[width=0.95\linewidth]{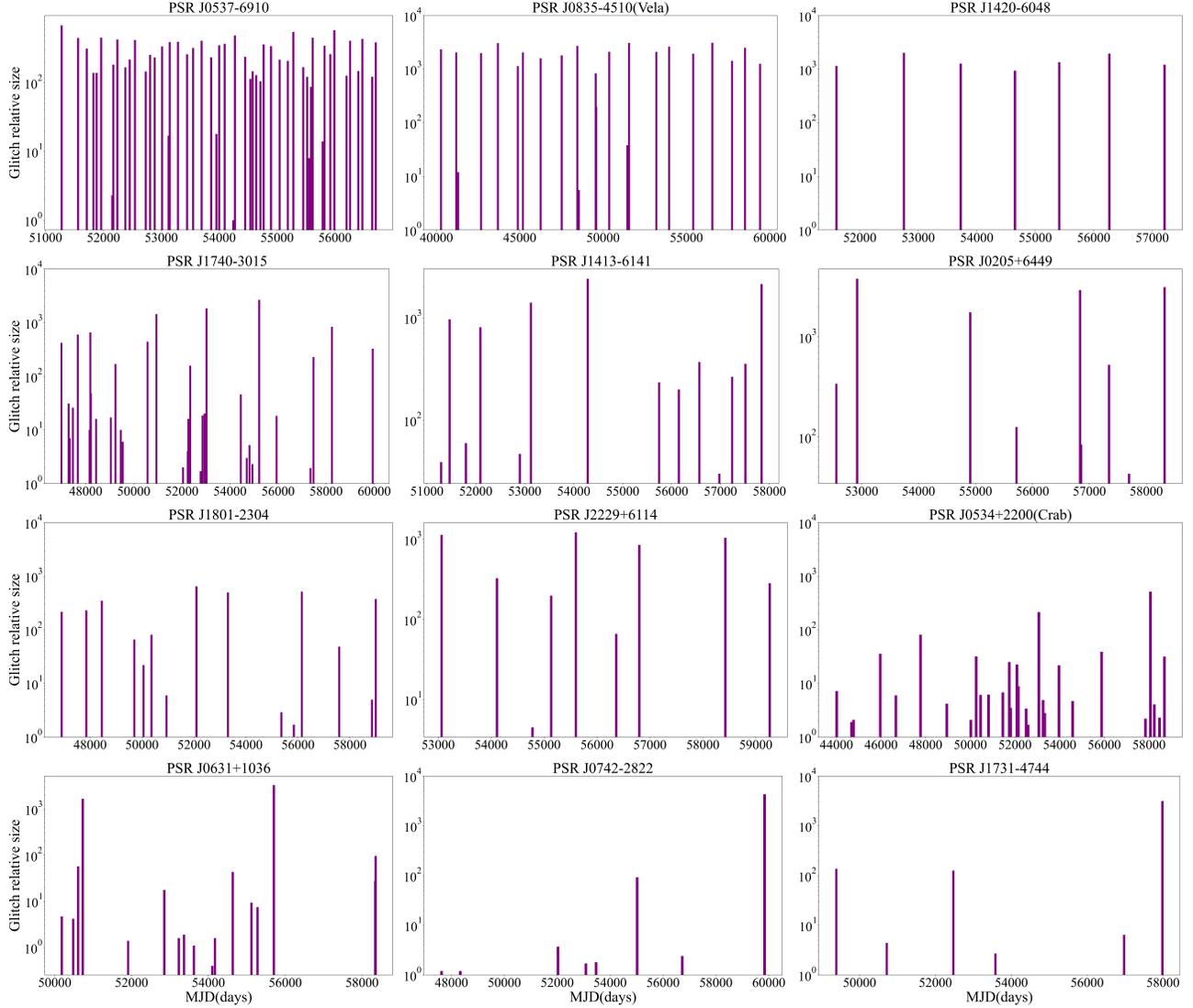}
	\caption{The bar chart illustrate the glitch sizes and occurrence times for 12 pulsars. In the top panel, a clear periodicity and similarity in glitch sizes are visually apparent, while the three pulsars in the second row show evident clustering distributions.  The panels in the third row fall somewhere between the two patterns, displaying weak clustering or subtle periodicity during certain time intervals.  The bottom panel features pulsars with significant glitches.}
	\label{fig:cluster}
\end{figure*}

\begin{figure*}
	\centering
	\includegraphics[width=0.95\linewidth]{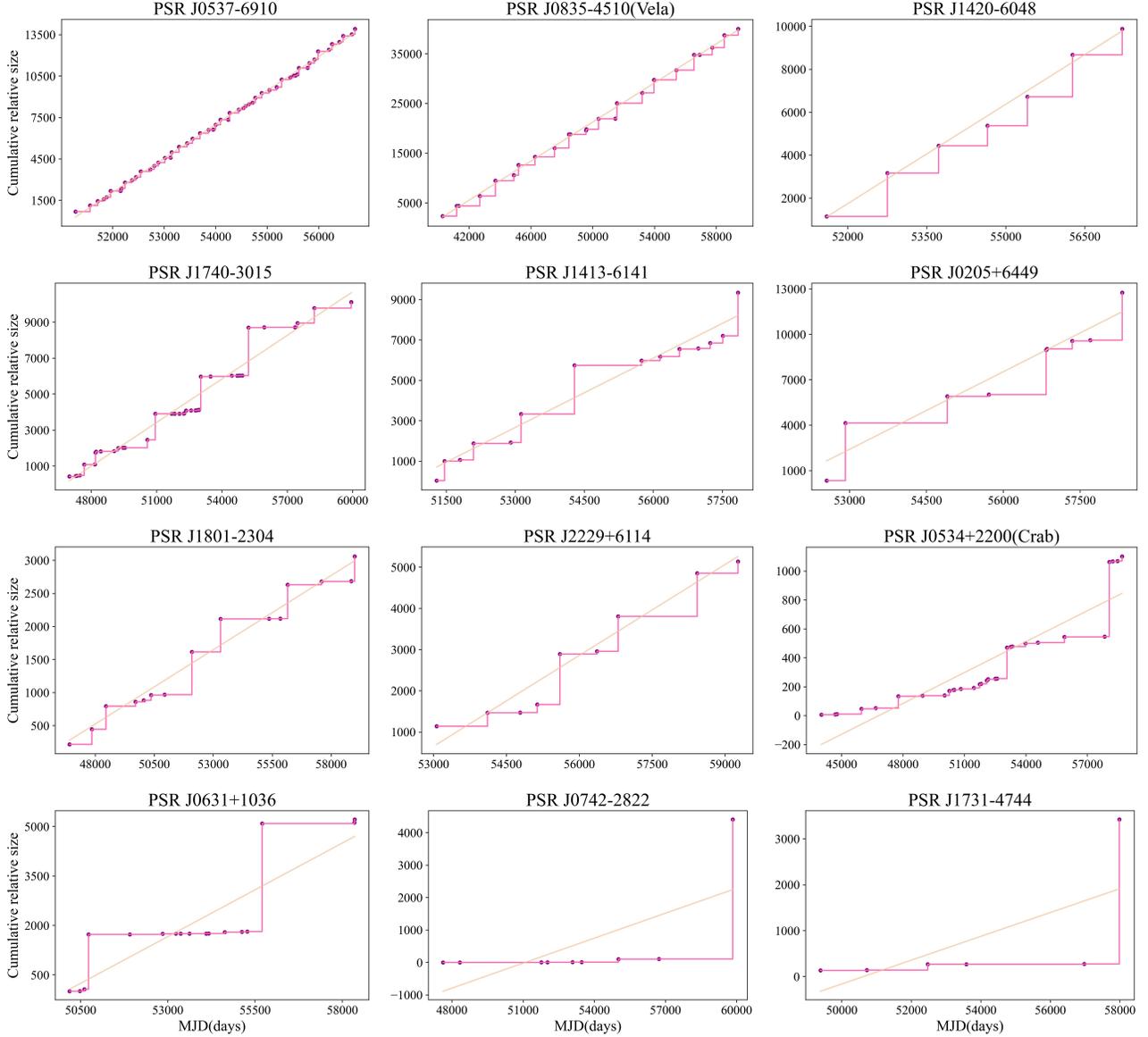}
	\caption{The pink lines represent the cumulative relative size of glitches over time for 12 pulsars, while the yellow line represents the linear fit. The top three pulsars exhibit uniform stability, whereas the second row of pulsars is characterized by large glitches.  The pulsars in the third row display a behavior that lies between these two patterns.  Conversely, the pulsars at the bottom are entirely dominated by one or two significant glitches.}
	\label{fig:cumulative}
\end{figure*}

\section{sample}
\label{sample}

The data used in this study are primarily drawn from the Australia Telescope National Facility (ATNF) pulsar catalogue\footnote{\url{https://www.atnf.csiro.au/research/pulsar/psrcat}} \citep{manchester2005australia} and the Jodrell Bank Observatory (JBO) pulsar glitch catalogue\footnote{\url{http://www.jb.man.ac.uk/pulsar/glitches.html}} \citep{espinoza2011study}
. Pulsar parameters are obtained from the ATNF catalogue, while the glitch data are sourced from the JBO glitch catalogue. Unless otherwise specified, relative glitch size is reported in units of $10^{-9}$ for $\Delta \nu/\nu$.
To ensure the statistical analysis remains meaningful and to minimize analytical errors as much as possible. Obviously, having a sufficient number of observed glitches and an appropriate density of glitch occurrences is crucial. Therefore, we do not consider pulsars with fewer than five glitches. 
Ultimately, we found that 32 pulsars have a sufficient number of glitches to meet this criterion. We here don't consider magnetars. Table \ref{tab:information} displays the detailed information of the selected 32 pulsars.

With the increasing accumulation of glitch observational data, conducting quantitative statistical analyses on individual pulsars, in addition to population-wide studies, plays a more critical role in understanding the glitch phenomenon. The frequent glitches of pulsars provide us with an opportunity to explore these events in greater detail.

For rare and sporadic events like glitches, the most straightforward observation involves visually inspecting glitch sizes and their time intervals. From these visual inspections, we initially discern four classes of glitching pulsars. 
In Figure \ref{fig:cluster}, We present 12 typical examples illustrating glitch sizes plotted against glitch epochs for each pulsar. The Vela pulsar, PSR J0537$ - $6910, and PSR J1420$ - $6048 belong to the first class, characterized by the production of glitches of similar sizes at reasonably regular time intervals \citep{mcculloch1987daily,eya2017angular}.  The second class of pulsars is characterized by the occurrence of multiple glitch events and appearance of a large glitch within a narrow time range, but with rare or even absent glitches occurring at other times, despite the presence of one or several significant glitches in the region. Examples of such pulsars are PSR J1740$ - $3015, PSR J1413$ - $6141, and PSR J0205$ + $6449. The third class of pulsars, represented by pulsars like PSR J1801$ - $2304, PSR J2229$ + $6114, and PSR J0534$ + $2200(Crab), exhibit less discernible regular traits based on our inspections. They may exhibit a mild degree of clustering during certain periods, but within this weak clustering, there is no clearly dominant glitch present. Lastly, the fourth category of pulsars is dominated entirely by one or two exceptionally large glitches. Examples include PSR J0631$ + $1036, PSR J0742$ - $2822, and PSR J1732$ - $4744. It's worth noting that there was a period of missing continuous observations for PSR J0537$ - $6910 and PSR J0534$ + $2200, with gaps of 584 days and three years, respectively. To ensure the inclusion of all available glitch data, we replaced the gaps in the data with the average glitch time interval.

Another narrative angle of view, the evolution of cumulative glitch sizes over time provides insights that may better illustrate the aforementioned perception of glitch behaviours, as depicted in Figure \ref{fig:cumulative}. In the figure, a yellow line represents the linear fit. Recent data indicates that glitches in the Vela pulsar exhibit quasi-periodicity, with a consistently stable cumulative size process. It can be observed that the purple points, indicating the glitches, closely align with the yellow fitted line. Similar behaviors are observed in pulsars such as PSR J0537$ - $6910 and PSR J1420$ - $6048.
However, in a distinct class of pulsars represented by pulsars like PSR J1740$ - $3015, PSR J1413$ - $6141, and PSR J0205$ + $6449, the cumulative size process is primarily influenced by a few significant glitches, despite the presence of numerous smaller glitches. Compared to the fit for the first class of pulsars (i.e., Vela-like pulsars), the deviation between the purple data points and the fitted line is more pronounced, indicating a larger discrepancy. The third class of pulsars, situated between these two modes, does not exhibit a clear overall trend in cumulative sizes of glitches. They may display weak clustering effects over shorter time intervals or exhibit subtle periodicity during specific periods. The final category includes pulsars such as PSR J0631$ + $1036, PSR J1732$ - $4744, and PSR J0742$ - $2822. In these cases, the cumulative angular momentum is predominantly contributed by major one or two glitches, while the numerous smaller glitches make negligible contributions to the overall angular momentum.
It is worth noting that, due to the intrinsic characteristics of glitch data and the limited number of observations available for individual pulsars, the distinction between the second and third categories remains somewhat tentative at this stage.

As an initial step in our quantitative analysis, we begin with a one-dimensional approach, focusing on the distributions of glitch sizes and waiting times for individual pulsars. For the 32 pulsars in our dataset, we test the cumulative distribution functions (CDFs) of both glitch sizes and waiting times using two null hypotheses: a normal distribution and a Poisson process with constant intensity ($ \lambda $). The Anderson-Darling (tail-weighted Cramer-von Mises) test is more sensitive than the commonly used Kolmogorov-Smirnov test. In Table \ref{tab:information}, we present the minimum significance levels at which the glitch sizes or waiting times can reject the two distributions.  Naturally, lower significance levels indicate greater deviation from the distribution. An asterisk ($ * $) denotes cases where the distribution cannot be rejected even at the 15\% significance level, suggesting that the data may potentially conform to the distribution based on the Anderson-Darling test. We find that for the Vela pulsar, PSR J0537$ - $6910, and PSR J1420$ - $6048, the distributions of glitch size and waiting time deviate from a homogeneous Poisson process at the 1\% significance level.  Conversely, PSR J1740$ - $3015, PSR J1413$ - $6141, PSR J0631$ + $1036, and PSR J0534$ + $2200 more robustly reject a normal distribution. Notably, PSR J1341$ - $6220 rejects both distributions. For the remaining pulsars, the Anderson-Darling test does not provide a clear distinction between the distributions to varying degrees.

In addition to AD testing on the 32 pulsars presented here, a substantial body of previous work has also focused on more precise modeling and analysis of glitch sizes and waiting times. \citet{melatos2008avalanche} investigated the statistical distributions of glitch sizes and waiting times in nine pulsars, discovering that seven of these pulsars exhibited a power-law distribution for glitch sizes and an exponential distribution for waiting times, while Vela and PSR J0537$ - $6910 were more accurately described by a Gaussian function. \citet{howitt2018nonparametric} further validated these findings using a nonparametric kernel density estimator; however, PSR J1341$ - $6220 displayed evidence of hybrid behavior. Furthermore, \citet{eya2017angular} determined that the glitch sizes of twelve pulsars conform to a normal distribution. However, it is undeniable that when conducting precise modeling of glitch data, a decrease in the number of glitches results in limited observational data, leading to reduced information extraction.  As a result, it becomes difficult to ascertain whether the observed distribution is a reflection of the intrinsic characteristics of the glitches or primarily a consequence of insufficient observational data, leading to the omission of critical details. 

In addition to analyzing the characteristics of one-dimensional data, understanding the relationship between glitch size and waiting time represents a particularly important aspect of studying discrete events such as glitches. The prevailing consensus in most models suggests that glitches result from the release of accumulated angular momentum within a neutron star, which accumulate at a rate determined by the spin-down rate \citep{haskell2015models, antonopoulou2022pulsar, antonelli2023insights}. If pressure were to be completely released at each glitch, there should be a certain correlation between glitch size and waiting time. Furthermore, based on the assumption that reaching a critical state is necessary for a glitch to occur, glitches should occur periodically with nearly equal sizes. While some pulsars exhibit quasi-periodicity, only PSR J0537$ - $6910 demonstrates a correlation between glitch size and the time to the next glitch \citep{middleditch2006predicting}. 

\begin{figure}
	\flushleft
	\includegraphics[width=0.9\linewidth]{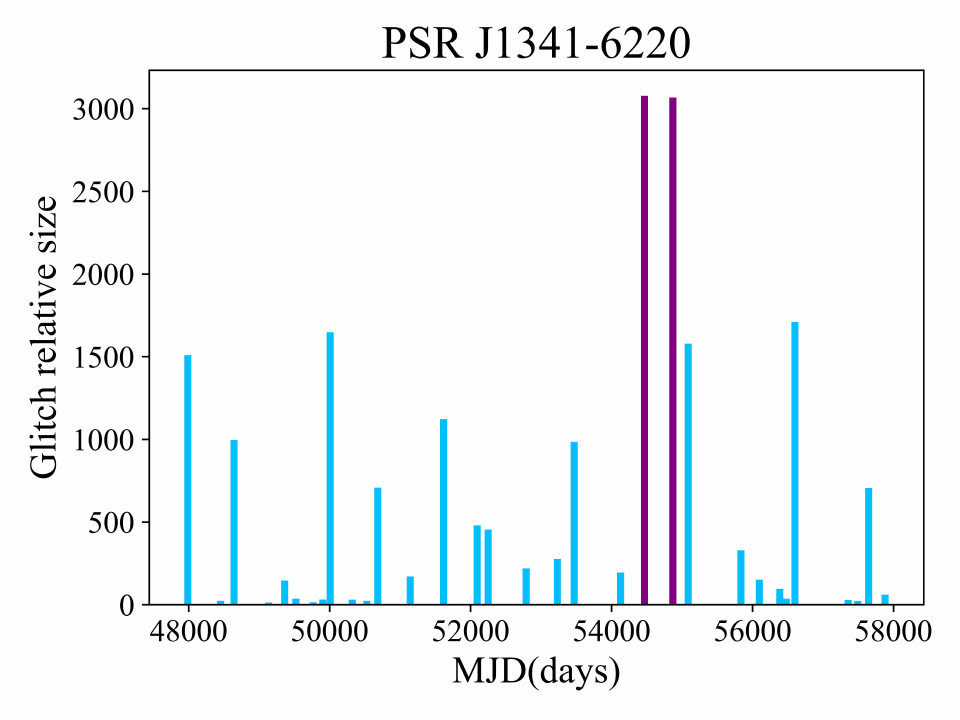}
	\caption{The distribution of glitch sizes over time for PSR J1341$ - $6220 is depicted. Notably, the consecutive 21st and 22nd glitches (highlighted in purple) indicate the occurrence of two significant glitches within a relatively short time frame.}
	\label{fig:PSR J1341-6220bar}
\end{figure}

Upon closer examination of the glitch data from the Vela pulsar, it is evident that four glitches exhibit $ \Delta \nu/\nu$ values of $10^{-9}$, specifically 12, 5.55, 38, and 0.4. These values are three orders of magnitude smaller than the other 20 glitches, which have $ \Delta \nu/\nu$ values of $10^{-6}$. Furthermore, it is noticed that the time intervals between these four glitches and nearby large glitches (with an average of 124.17 days) are significantly shorter than the overall average time intervals (with an average of 956.85 days). The same phenomenon is observed in PSR J0537$ - $6910. This suggests that even in pulsars exhibiting good quasi-periodicity, glitch sizes can vary significantly.

In addition, there is a perplexing observational fact regarding PSR J1341$ - $6220 (B1338$ - $62). Among the 35 glitches observed, the 21st and 22nd glitches (highlighted in purple) stand out as highly unusual. The relative sizes $ \Delta \nu/\nu$ of these two glitches were 3078.2 and 3066.7, significantly higher than other glitches, and they occurred in close proximity, with a time interval of 402 days, as shown in Figure \ref{fig:PSR J1341-6220bar}. Based on mainstream models, explaining how PSR J1341$ - $6220 could accumulate and release such enormous pressure in such a short time presents a significant challenge.

Various statistical analyses and tests offer valuable insights into different aspects of the data; however, in the current context of limited samples, accurately modeling the glitch sizes and waiting times of different pulsars remains a challenge. Drawing from the four broad classifications presented in Figures \ref{fig:cluster} and \ref{fig:cumulative}, it is crucial to establish a cohesive understanding of the quasi-periodicity observed in pulsars, the potential for short-term clustering, and the occurrence of one or two large glitches. Our study seeks to provide a unified interpretation of the glitch phenomena exhibited by various pulsars.

To more effectively quantify and highlight the distinctions among the four types of pulsars, we adopt an empirical approach by defining a dimensionless quantity—the relative area error—as a means of assessment. In Figure \ref{fig:cumulative}, we present the cumulative glitch size for different pulsars as a function of time, with purple points marking individual glitch events and the yellow solid line representing the linear fit. For each purple point, horizontal and vertical lines are extended to intersect with the yellow fit line, forming a triangle. The areas of these triangles are calculated and summed to define an error parameter, $ \Delta S $. Empirically, a larger value of $ \Delta S $ indicates greater deviation from the linear trend. Due to variations in sample size, observation times, and glitch magnitudes among pulsars, normalization of this parameter is necessary for meaningful comparisons. Thus, we use the area of the large triangle formed by the sizes of all glitches and the total time span of different pulsars as a reference value to calculate the relative value of $ \Delta S $. The final column of Table \ref{tab:information} represents  the normalized error parameter for each pulsar.
Combining the data from Table \ref{tab:information} and the classification in Figure \ref{fig:cumulative}, we observed that for the first type of pulsars, $ \Delta S <1$, while for the second type, $ \Delta S \sim 5-8$. The third type of pulsars exhibits $ \Delta S $ values that fall between those of the first and second types. For pulsars with exceptionally large glitches, $ \Delta S $ shows both higher values and greater uncertainty.

\section{Glitch time series and clusters}
\label{clusters}
To bridge these differences in glitches of the pulsars, we here attempt to reveal the presence of long-term clustering phenomenon. Building upon this clustering concept, we can analyze the overall representation of glitches. This perspective differs from the traditional treatment of glitch individuals, where each glitch is viewed independently. Instead, it suggests that we should reevaluate either periodicity or randomness in glitch manifestations from a new standpoint.

\begin{figure*}
\centering
\includegraphics[width=0.95\linewidth]{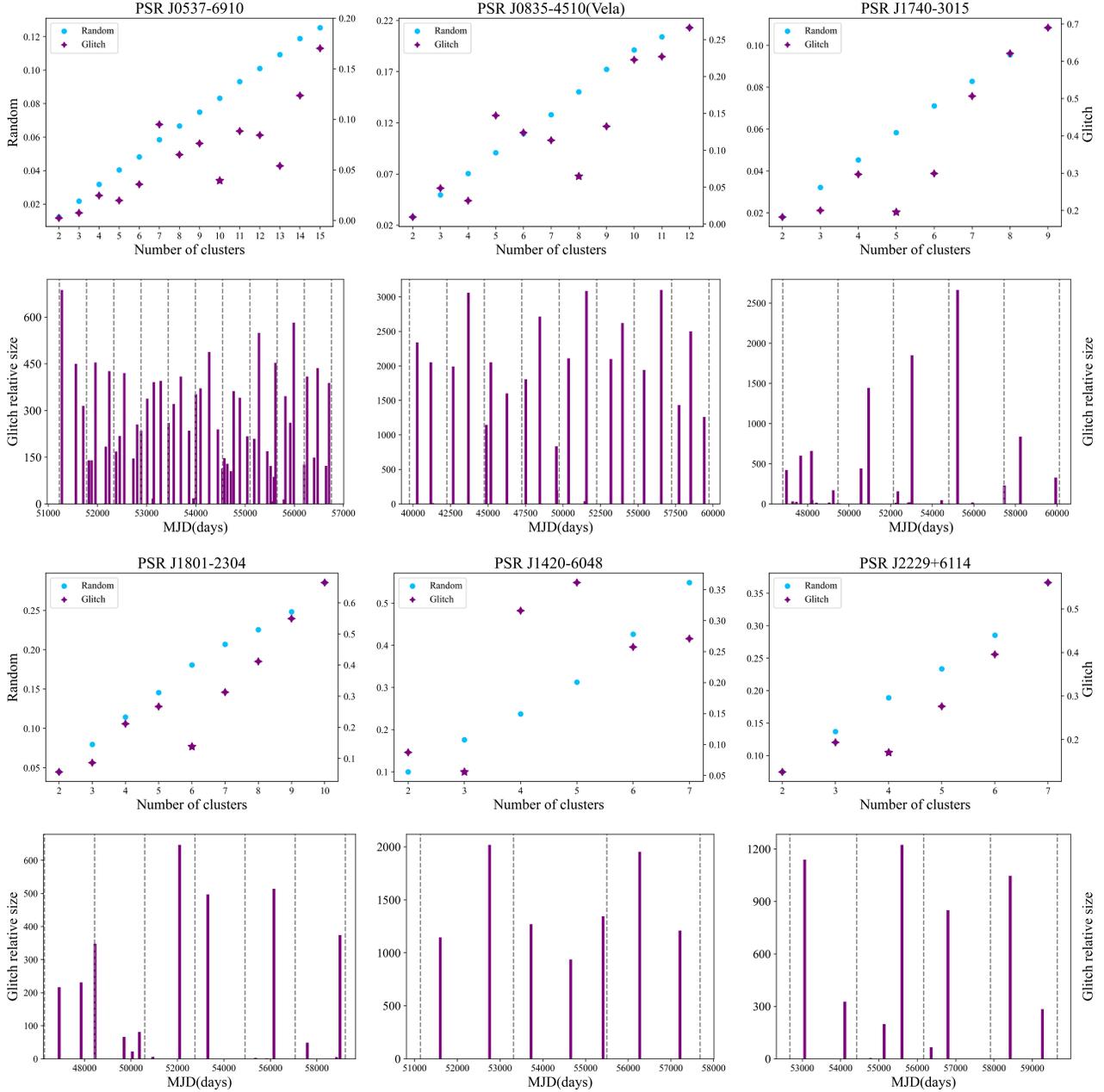}
\caption{The plot illustrates changes in the coefficient of variation for six pulsars under different grouping scenarios.  Comparing them with the coefficient of variation of random sequences (in blue), we mark the optimal groupings with purple pentagrams.  Using the calculated average glitch cluster periods, we annotate the distribution of glitch sizes over time with equidistant dashed lines for different pulsars.}
\label{fig:clustering}
\end{figure*}

To quantitatively analyze the overall behavior of glitches over longer timescales, the primary approach is to treat multiple glitches as a single glitch cluster. The key challenge lies in grouping glitches without altering their sizes or occurrence times, while ensuring that the resulting clusters accurately capture the intrinsic properties of the pulsar's glitching behavior.

For the glitch data of a given pulsar, we first consider the total sum of glitch sizes within each group under different numbers of groups. Building upon this, we compute the coefficient of variation of the total sum of glitch sizes across all groups to assess the effectiveness and validity of the grouping. In statistics, the coefficient of variation (CV) is a normalized measure of the dispersion of a probability distribution, primarily used to compare the degree of variability between different data samples. It is defined as the ratio of the standard deviation to the mean. We observe variations in the coefficient of variation for different pulsars under different numbers of groups, as depicted in Figure \ref{fig:clustering}.

To better capture the distinctive features of glitch data and its intrinsic grouping properties, we perform a comparative analysis using random datasets.
Taking the Vela pulsar as an example, we generated 2000 random sequences, each comprising 24 data points. Subsequently, we sequentially grouped these 2000 sets of random sequences from 2 groups to 24 groups. Finally, we compared the average of these 2000 grouping results with the glitch data, as depicted in Figure \ref{fig:clustering}. 
In the plots, the blue data points represent the coefficient of variation of the random sequences under different groupings, while the purple data points denote the glitch data. Independent of sample size and number of groups, we observe that the coefficient of variation for random sequences consistently increases as the number of groups increases.
However, we found that for some pulsars, the coefficient of variation exhibits multiple minima, and the numbers of groups corresponding to these minima are in multiples. For instance, this behavior is observed in Vela, PSR J0537$ - $6910, and PSR J1740$ - $3015.
To mitigate the influence of varying value ranges in the random sequences, we normalized the coefficient of variation for the maximum number of groups (24 groups). To identify the optimal grouping and eliminate subjective influences in the selection process, we calculated the relative discrepancy between the two datasets (glitch data and random sequences) and used this relative discrepancy as the criterion for determining the best grouping. The optimal grouping is marked with purple pentagrams. The specific grouping data for the Vela pulsar is presented in Table \ref{table:vela}.

This approach allows us to visually highlight the most significant differences between the glitch data and the optimal grouping method compared to the random sequences. In other words, this method helps identify the intrinsic grouping properties inherent in the glitch data itself. To provide a clearer demonstration of the grouping results, we only present the initial part of the variations for some pulsars, where the changes are more pronounced and apparent, especially when the number of groups is relatively small.

This grouping method focuses on capturing the long-term periodicity in glitch data, which means that the coefficient of variation for glitch clusters must be lower than that of random sequences for the grouping to be considered valid. Additionally, we exclude cases where the optimal grouping number is 2, as such groupings not only fail to yield meaningful results but also introduce significant errors in determining the cluster period. After applying identical grouping and coefficient of variation comparisons across 32 pulsars, we find that only 17 pulsars meet the above criteria, corresponding to the first 17 pulsars shown in Table \ref{tab:information}.

\begin{deluxetable*}{cccccccccccc}
	\tablecaption{The detailed grouping data for the Vela pulsar.\label{table:vela}}
	\tablehead{
		$N$ & CV\_glitch & CV\_random & $ \delta $ & 	$N$ & CV\_glitch & CV\_random & $ \delta $ &	$N$ & CV\_glitch & CV\_random & $ \delta $}
	\startdata
	2 & 0.010  & 0.028  & -0.682  & 10 & 0.223  & 0.193  & 0.091  & 18 & 0.283  & 0.358  & -0.254   \\ 
	3 & 0.049  & 0.050  & -0.087  & 11 & 0.227  & 0.203  & 0.055  & 19 & 0.305  & 0.403  & -0.284   \\ 
	4 & 0.032  & 0.070  & -0.574  & 12 & 0.266  & 0.214  & 0.174  & 20 & 0.388  & 0.446  & -0.179   \\ 
	5 & 0.147  & 0.090  & 0.539  & 13 & 0.297  & 0.223  & 0.256  & 21 & 0.456  & 0.484  & -0.111   \\ 
	6 & 0.124  & 0.110  & 0.063  & 14 & 0.242  & 0.235  & -0.027  & 22 & 0.503  & 0.518  & -0.081   \\ 
	7 & 0.114  & 0.129  & -0.171  & 15 & 0.240  & 0.253  & -0.107  & 23 & 0.555  & 0.545  & -0.038   \\ 
	8 & 0.065  & 0.149  & -0.588  & 16 & 0.255  & 0.281  & -0.143  & 24 & 0.604  & 0.571  & 0.000   \\ 
	9 & 0.132  & 0.172  & -0.273  & 17 & 0.264  & 0.316  & -0.209  & ~ & ~ & ~ &   \\ 
	\enddata
	\tablecomments{$N$ represents the number of groups, ranging from 2 to 24.  CV\_glitch and CV\_random indicate the coefficients of variation for the glitch data and the random sequence, respectively, under each grouping.  The relative difference, $ \delta $, is calculated as $ \delta $  = ( {CV\_glitch} $ - $ {CV\_random} ) / {CV\_random}.  Excluding the case where $N=2$, the smallest $ \delta $ value is -0.588, corresponding to a grouping of 8.}
\end{deluxetable*}

Once the optimal grouping is determined, we can establish the period of the glitch cluster.
Since we can only determine the total time span of the glitches, which is the time difference between the first and the last glitch, but lack information on the actual observation duration, the main source of error in determining the glitch cluster period arises from the unknown time before the first glitch and after the last glitch. 
To address this uncertainty, we utilized the average interval between glitches to make a reasonable estimation. Consequently, we calculated the average period for each glitch cluster using the formula
\begin{displaymath}
\bar{P}_C=\frac{T}{N_C-{1}/{\bar{N}_{gC}}} ,
\end{displaymath}
Here, $ T $ represents the observational time span, i.e., the time difference between the occurrence of the last glitch and the first glitch. $ N_C $ denotes the optimal number of groups we determined, while $ \bar{N}_{gC} $ represents the average number of glitches within each group, i.e., the total number of glitches divided by the number of groups. We do not directly use $ \bar{P}_C = T/N_C $, as this essentially applies a correction to the time span.
Based on the calculated average periods of glitch clusters, we marked the glitch distribution bar chart with equidistant intervals and found that the results of the majority of pulsars, determined numerically and by equidistant grouping, are highly consistent. However, PSR J1341$ - $6220 stood out as an exception, where the distribution of some small glitches did not precisely match the numerical results under equal-interval grouping. Therefore, we believe that a non-equidistant distribution is more realistic, implying that the cluster period of each pulsar is not a fixed value.

While the statistical and grouping methods mentioned in this study may seem somewhat simple and straightforward from a technical perspective, they prove advantageous in capturing the intrinsic properties of glitch clusters, especially considering the diverse manifestations of glitches and the limited amount of data available. Many statistical methods struggle to adapt effectively to such scenarios. Therefore, in the current context, we believe that the method proposed in this paper has distinct advantages in reflecting the inherent characteristics of glitch data itself.

For the 17 pulsars that meet the above conditions, the total observational timespan is typically limited to no more than five or six decades, or even less. Combined with the grouping procedure, this results in a dataset confined to a relatively narrow range, making it difficult to explore larger-scale trends or incorporate additional data. To address this limitation, We utilize ten pulsars characterized by one or two exceptionally large glitches, classified as the fourth category in our earlier pulsar classification. Examples include PSR J0631$ + $1036, PSR J0742$ - $2822, and PSR J1731$ - $4744, which serve as potential supplementary samples. Given the current characteristics of the glitch data, we tentatively treat the intervals between these large glitches as the periods of glitch clusters.

The five excluded pulsars are PSR J0534$ + $2200,  PSR J1841$ - $1744, PSR J2021$ + $3651, PSR J1902$ + $0612, and PSR J2225$ + $6535. The first three were excluded because they did not meet the strict grouping criteria and lacked exceptionally large glitches that could be used as supplementary data. The last two pulsars were excluded due to their characteristic ages reaching the million-year scale, which introduces significant uncertainties.

\begin{figure}[ht!]
	\flushleft
	\hspace{1mm}
	\includegraphics[width=0.95\linewidth]{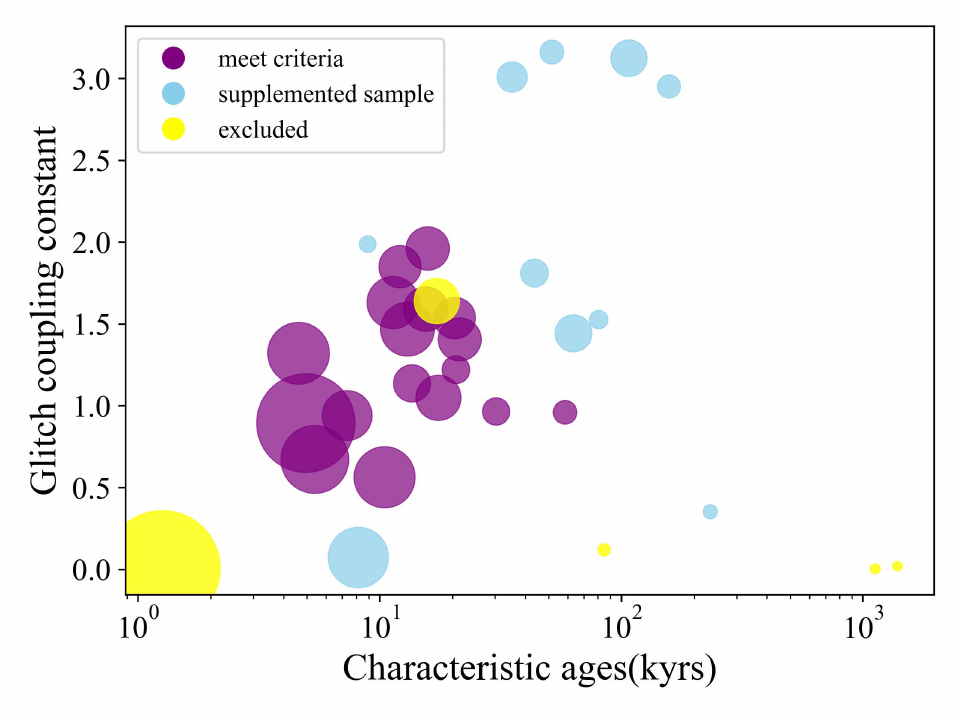}
	\caption{The figure depicts the relationship between glitch coupling constant $G$ and characteristic age for 32 pulsars, with the size of the circles representing the magnitude of the frequency derivative (non-linear scale). We only consider pulsars with more than 5 glitches. Seventeen purple circles represent the pulsars that met the criteria during the grouping process, ten blue circles represent the pulsars included as supplementary data, and five yellow circles denote the pulsars that were excluded.}
	\label{fig:sample}
\end{figure}

Figure \ref{fig:sample} shows the relationship between the glitch coupling constant and characteristic age of pulsars with five and more than five glitches. The size of each point represents the magnitude of the first derivative of the pulsar's frequency. 
We have marked the 17 pulsars that meet criteria in purple, the 10 pulsars with supplemented data in blue, and the 5 pulsars that were excluded in yellow.

Furthermore, for some pulsars, based on the characteristics of the glitch data, we considered the possibility that the first or last one or two glitches might not have completed a full period. However, upon analysis, we found that when the number of glitches is relatively large (N $>10  $), within the entire observational time span ($ 10^{3} \sim 10^{4} $ days), the time span of one or two glitches has little impact on determining the glitch cluster period.

\section{statistical analysis of glitch clusters}
\label{statistical}
\begin{figure}
	\flushleft
	\includegraphics[width=0.95\linewidth]{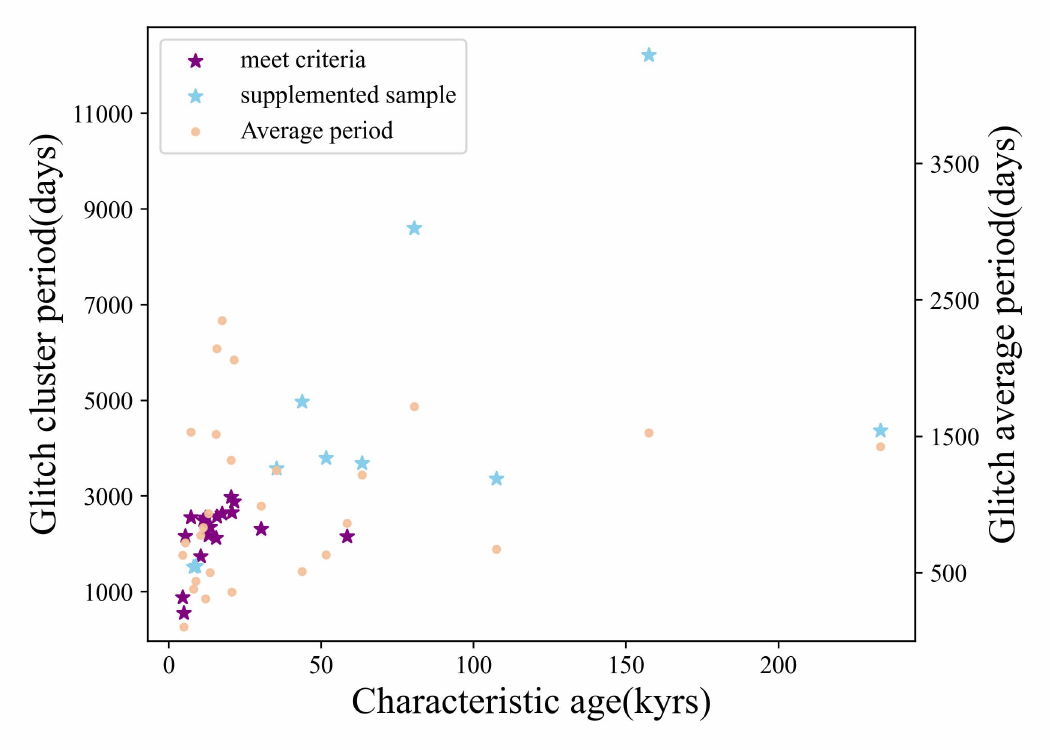}
	\caption{The purple pentagrams correspond to the 17 pulsars that strictly adhere to the grouping criteria, while the blue pentagrams represent 10 additional pulsars included as supplementary samples. For comparison, the yellow points indicate the ungrouped average glitch intervals.}
	\label{fig:period}
\end{figure}

\begin{figure*}
	\centering
	\includegraphics[width=0.95\linewidth]{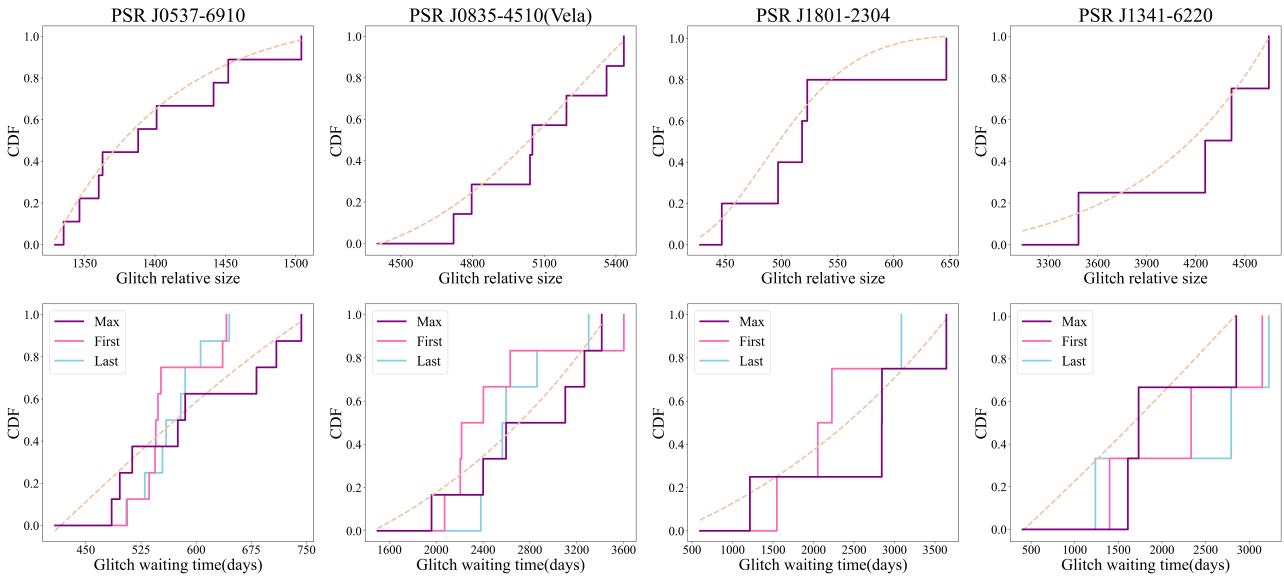}
	\caption{The top row illustrates the CDF of glitch sizes for four pulsars, analyzed using glitch clusters instead of individual glitches. Below, the CDF distribution of waiting times between glitch clusters is displayed. We mark three specific time reference points: the first, last, and largest glitches within each glitch cluster. These points aid in accurately determining the time intervals between adjacent glitch clusters. The yellow dashed line represents the Gaussian fit. For the waiting time CDFs, we only display the fitting with the largest glitch as the reference point within each glitch cluster.}
	\label{fig:cdf}
\end{figure*}
\begin{deluxetable}{ccccc}
	\tablecaption{Correlation analysis between the  characteristic age $ \tau_{c} $ and both average period and cluster period. The second and third column indicates the Spearman's correlation and  its p-value respectively. The fourth and fifth column show Kendall-tau correlation coefficient and its p-value respectively.\label{table:period}}
	\tablehead{
		\colhead{period-versus- $ \tau_{c} $} & \colhead{$r_s$} & \colhead{$r_s(p$-value $)$} & \colhead{$\tau$} & \colhead{$\tau(p$-value $)$ }}
	\startdata
	average period & 0.221  & 0.268  & 0.282  & 0.040  \\
	cluster period & 0.643  & $2.96 \times 10^{-4}$ & 0.670  & $9.5 \times 10^{-8}$ \\
	\enddata
	
\end{deluxetable}
Using the aforementioned method, we determined the glitch cluster periods for 27 pulsars.  We first analyzed the relationship between the glitch cluster periods and spin parameters across different pulsars.  
Notably, we observed a significant linear relationship between the glitch cluster period and the characteristic age of the pulsars, as illustrated in Figure \ref{fig:period}. The purple points denote pulsars that strictly adhere to the grouping criteria, whereas the blue points represent supplementary pulsars. Incorporating large glitches as supplementary data may, to some extent, reflect underlying trends in the evolution of glitch clusters. As a comparison, the yellow data points represent the average time intervals between glitches before any grouping was applied. To provide a more quantitative assessment of this correlation, we applied Spearman's rank correlation test and the Kendall–Tau test. The results are summarized in Table \ref{table:period}. However, PSR J1801$ + $2304 shows a significant deviation from the established linear trend. This deviation may be due to uncertainties in the characteristic age derived from the magnetic dipole braking model, or possibly due to undetected glitches.

This linear relationship is consistent with our understanding of glitches: as the characteristic age increases, pulsars require more time to complete a cycle. 
This also implies that even in older pulsars, large glitches may still occur. For instance, PSR J2225$ + $6535, with a characteristic age of approximately 1.1 million years and a relatively low $G$ of around $\sim 4\times10^{-5}$, exhibited a significant large glitch, as indicated in Table \ref{PSR J2225+6535}. However, it is worth noting that the accuracy of this linear relationship heavily depends on pulsars with larger characteristic ages, besides  the uncertainties associated with the characteristic ages calculated from the magnetic dipole model. This reliance stems not only from the scarcity of pulsars with sufficiently large characteristic ages suitable for analysis, but also from the fact that pulsars with larger characteristic ages exhibit fewer glitch occurrences and longer glitch cluster periods. Therefore, longer observational periods and larger samples of pulsar glitch data are required to better examine and refine this linear relationship. If this linear relationship is corroborated by further data, it may provide a novel avenue for estimating pulsar ages through the investigation of glitches as a dynamical phenomenon.

\begin{deluxetable}{cccc}
	\tablecaption{Glitch data for PSR J2225$ + $6535.\label{PSR J2225+6535}}
	\tablehead{
		\colhead{J-name} & \colhead{No.(Glt's)} & \colhead{MJD(day)} & \colhead{$\Delta\nu / \nu$($10^{-9}$)}}
	\startdata
	2225$ + $6535 & 1 & 43072 & 1707 \\
	2225$ + $6535 & 2 & 51900 & 0.14 \\
	2225$ + $6535 & 3 & 52950 & 0.08 \\
	2225$ + $6535 & 4 & 53434 & 0.2  \\
	2225$ + $6535 & 5 & 54266 & 0.4 \\	
	\enddata
\end{deluxetable}

On the basis of our definition of glitch cluster, we treat cluster as individual entities for statistical analysis. We generate cumulative distribution functions (CDFs) of cluster sizes and times between clusters, comparing them with typical distribution patterns. Figure \ref{fig:cdf} shows the CDFs of sizes for four representative pulsars. The size of a cluster is simply the sum of each glitch size within the cluster. Due to constraints in available glitch observational data, the number of groupings is generally limited. It is important to note that in our plots, glitch sizes do not start from zero due to the glitch cluster assumption. Since there are two or more glitches in a cluster, the determination of time intervals between consecutive clusters requires the appointment of specific reference points. Here, we choose a few time intervals: between the first glitches, the last glitches, and the largest glitches within each pair of clusters. The CDFs of ``interval time" are respectively shown in Figure \ref{fig:cdf}.

An analysis of the cumulative distribution functions (CDFs) of glitch sizes and waiting times for different pulsars reveals distinct distribution patterns, including Gaussian, power-law, or exponential distributions, indicating significant heterogeneity among pulsars \citep{fuentes2019glitch}. This diversity has posed challenges to our understanding of the glitch phenomena. In an ideal scenario, as predicted by theoretical models, the sum of glitch sizes for each period should be equal, resulting in the convergence of CDFs into a vertical line. However, this representation is unattainable with realistic glitch data. Nevertheless, when considering glitch clusters, a consistent pattern emerges in the CDFs after grouping glitches together, as depicted in Figure \ref{fig:cdf}. We find that both the size and interval time CDFs can be well-fitted by Gaussian functions, especially the size CDFs. In the figure, the fitting is represented by yellow dashed lines. Of course, the interval time CDFs, influenced by the subjective factor of reference point selection and the overall linear relationship, may exhibit somewhat shift from  Gaussian characteristics. For the fitting of the waiting time CDFs, we only display the fitting with the largest glitch as the reference point within each glitch cluster. Similarly, we performed the AD test on the grouped glitch ensemble data. Given the constraints of the optimal number of glitch groups, we focused on eight pulsars that at least five groups. The final results demonstrate that, within the framework of the AD test, these eight pulsars not only display size and waiting time distributions consistent with normality, but they also increase the likelihood of rejecting a homogeneous Poisson distribution to some extent.

\section{discussions and Conclusions}
\label{Conclusion}
Analyzing the glitch coupling constant $G$ over time is crucial for understanding the superfluid reservoir \citep{link1999pulsar,ho2015pinning}.
According to the definition of $ G $, it not only reflects the acceleration effects induced by glitches on the long-term spin evolution of pulsars during the observation period, but its greater significance lies in its capacity, within the framework of the superfluid model, to quantify the ratio of the moment of inertia of the superfluid that dominate and contribute to glitches to the overall moment of inertia of the neutron star. 

\begin{figure}
	\flushleft
	\includegraphics[width=0.95\linewidth]{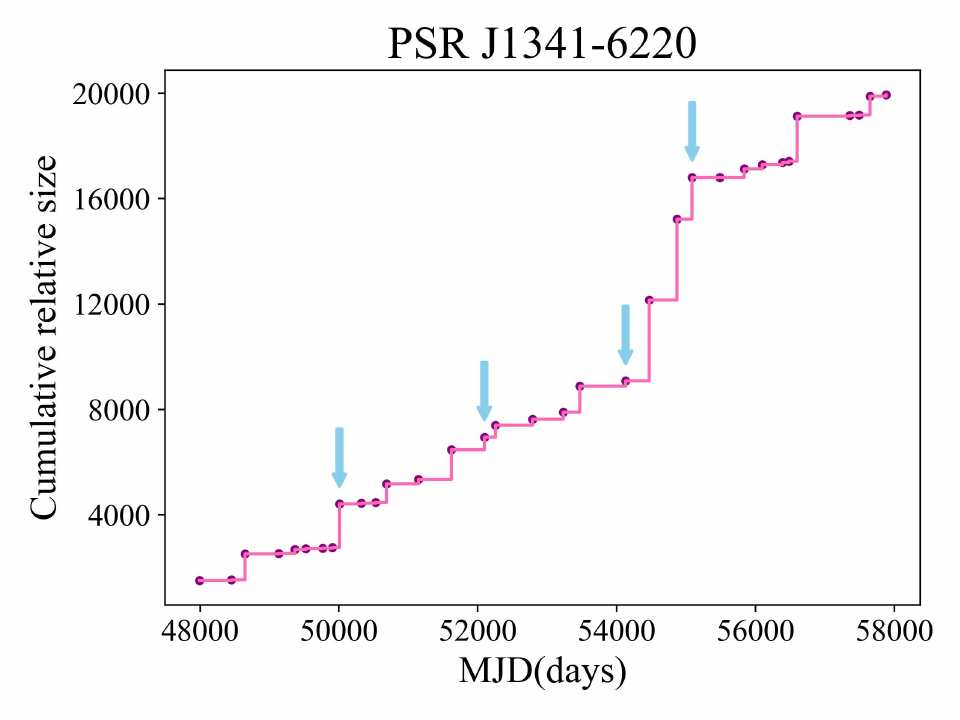}
	\caption{The temporal variation of accumulated angular momentum in PSR J1341$ - $6220 is illustrated, with four blue arrows delineating five distinct phases.}
	\label{fig:6220size}
\end{figure}

It is essential to recognize that $ G $ reflects the average effect derived from the total observation time span. For pulsars such as Vela, which display good quasi-periodicity, $ G $ is unlikely to exhibit significant variation over shorter timescales relative to the overall observation period. However, this may not be the case for other pulsars. As illustrated in Figure \ref{fig:cumulative}, the manifestations of glitches vary significantly among these pulsars. 
Additionally, PSR J1341$ - $6220 represents an even more extreme case, characterized by the adjacent occurrence of its two largest glitches, which leads to a substantial accumulation and release of angular momentum within a short timeframe. Considering the phenomenon from the perspective that each glitch occurrence completely empties the angular momentum of the superfluid reservoir is clearly unreasonable. A detailed analysis of the glitch data for PSR J1341$ - $6220 reveals notable fluctuations in $G$ over shorter timescales. To provide a clearer illustration of this variation, we have divided the glitch data for PSR J1341$ - $6220 into five phases, as shown in Figure \ref{fig:6220size}.  The boundaries are marked by four blue arrows, while the specific data are presented in Table \ref{table:6220}. We find that the $ G $ for the first three phases are 1.28\%, 1.08\%, and 0.93\%, respectively. This indicates that, over a comparable timescale of approximately 2000 days, $ G $ decreases steadily. However, in the fourth phase, which spans approximately 1000 days, the influence of two significant glitches results in an abrupt increase in $ G $ to 7.13\%, which is markedly higher than the expectations of the superfluid model.

\begin{deluxetable}{cccc}
	\tablecaption{The glitch data for PSR J1341$ - $6220 across different phases.\label{table:6220}}
	\tablehead{	\colhead{phase} & \colhead{No.(Glt's)} & \colhead{MJD(day)} & \colhead{$ G $(\%)}}
	\startdata
	1 & $ 1-9 $ & $ 47989-50008 $ & 1.28  \\ 
	2 & $ 9-15 $ & $ 50008-52093 $ & 1.08  \\ 
	3 & $ 15-20 $ & $ 52093-54128 $ & 0.93  \\ 
	4 & $ 20-23 $ & $ 54128-55088 $ & 7.13  \\ 
	5 & $ 23-33 $ & $ 55088-57880 $ & 0.99  \\ 
	\enddata
\end{deluxetable}
Therefore, we can reasonably infer that, in the case of PSR J1341$ - $6220, over an extended timescale ($ \sim $6000 days), although multiple glitches facilitate the release of this accumulated angular momentum, not every glitch leads to a complete depletion. It is probable that some residual angular momentum remains unspent, resulting in an increasing accumulation over time, which ultimately culminates in the occurrence of two significant glitches. A rough approximation indicates that the average $ G $ prior to the occurrence of large glitches is 1.09\%.  If glitches continue to release angular momentum at this rate (i.e., assuming $ G $ remains constant), by MJD 55088, the cumulative relative size of the glitches would be approximately 10500, while the actual observed value is around 16800.  This discrepancy suggests that over an extended timescale, PSR J1341-6220 has a sufficiently large superfluid reservoir to accumulate angular momentum, which could generate at least 6000 in glitch size (assuming the superfluid reservoir is fully depleted following the occurrence of these two substantial glitches). If our observation were to conclude before MJD 54128—during a phase of average $ G $ of 1.09\% —we would remain unaware that the pulsar is continuously accumulating and storing residual angular momentum. Thus, a simple estimate can be made by averaging the glitch that occurred in the fourth phase over the first four time intervals. This suggests that approximately 60\% of the angular momentum accumulated in the superfluid reservoir is released, with up to 40\% retained. Calculating the ratio of the slope over the first four phases to that over all five phases suggests that the actual moment of inertia of the superfluid vortices involved in glitches may range between 1 and 1.16 times the observed value of $ G $.

It is important to acknowledge that pulsars have the capacity and scale to accommodate the long-term accumulation of excess angular momentum, even if such accumulation may not be evident on shorter timescales. In summary, the active superfluid reservoir represented by $ G $ and the maximum superfluid reservoir that a neutron star can sustain are distinct concepts. The use of adjacent glitches to calculate $ G $ may yield values that vary significantly \citep{eya2017angular}.
From the perspective of glitch clustering, we can establish the minimum observational timescale necessary for the meaningful definition of $ G $ for each pulsar. This indicates that the definition of $ G $ is valid only beyond this specific timescale.

The Crab Pulsar, being very young and having a significantly lower glitch $G$ by about three orders of magnitude ($10^{-5}$) compared to other pulsars, presents an anomalous case \citep{fuentes2017glitch}. The internal temperature of the Crab Pulsar is relatively high, and the nuclear matter hasn't completely formed a superfluid state \citep{yakovlev2004neutron,andersson2021superfluid}. Therefore, the Crab Pulsar exhibits distinct slow rise and permanent deviations in its observational behavior, and over time, its glitch sizes not only maintain randomness but also show a tendency towards larger glitch amplitudes \citep{shaw2018largest,ge2020discovery}. Therefore, the new methods are required to determine the glitch cluster period for this specific pulsar.

However, the introduction of nondissipative
entrainment effects within the crust has revealed that these effects require a greater proportion of superfluid neutrons—specifically, a ratio of 4.3—for glitches to occur\citep{andersson2012pulsar,chamel2013crustal}. Given the pressure transition between the crust and core, along with the uncertainties regarding the properties of superfluids at high densities, further investigation is necessary to ascertain whether the inner crust can adequately store angular momentum and whether the superfluid in the core contributes to glitch\citep{piekarewicz2014pulsar,steiner2015using,li2016structures}.

In Figures \ref{fig:cluster} and \ref{fig:cumulative}, we broadly categorized pulsars into four types, with the 17 pulsars that strictly meet the grouping criteria concentrated in the central region of Figure \ref{fig:period}. The supplementary samples, in contrast, are either older or exhibit higher or lower values of $ G $. Thus, it is likely that variations in the internal states of neutron stars contribute to the existence of these four categories, which are organized based on the characteristics of glitch manifestations within the broader neutron star population.
Following a glitch, a recovery process may occur that lasts from several dozen to hundreds of days. While numerous theoretical models have been proposed to characterize this recovery process, it remains uncertain whether the superfluid vortices involved in the glitch fully recover prior to the next occurrence \citep{alpar1993postglitch}. Detailed observations of changes in the first and second derivatives of frequency, along with short-term braking indices following a glitch, could help elucidate the ratio of release to recovery \citep{lyne1996very,espinoza2017new,akbal2017nonlinear}. Furthermore, these observations may provide insight into the mechanisms by which glitch behaviors evolve across different pulsar populations.

We studied the glitching behaviour in pulsars by virtues of glitch clusters instead of individual glitches. Our analysis was restricted to the 27 pulsars that have frequent glitched. Our main conclusions are as follows:
\begin{enumerate}
	\item We recognize the temporal cluster of all pulsar glitches in our sample, even for Vela and PSR J0537-6910. We determine the number of glitch clusters by minimizing the coefficient of variation of group sizes. Those results pass the comparative test with random sequences.
	\item Glitch clusters exhibit quasi-periodic behaviour. This suggest that the glitches in a cluster may draw their angular momentum from a common reservoir. However, understanding the irregular emergence of glitches requires further investigation, treating a glitch cluster akin to a seismically active period.
	\item Glitches clusters have approximately equal sizes for every object besides quasi-periodicity. Therefore, the distributions of both their sizes and interval times are best fitted by Gaussian for all samples, indicating well-defined scales in cluster level. All pulsars are unified into Gaussian distributions, which is differs from a variety of best-fitting distributions for individual glitches. Glitch clusters are correlated and events and glitched uncorrelated ones. Combine into the second item, we speculate that the reservoir is sufficiently accumulated to a critical point, but the angular momentum is not release all at once, multiple releases become a active period in a pulsar glitches.
	\item In general, glitch activity is liner dependence of spin-down rate \citep{basu2022jodrell}, with more recent research suggesting that a quadratic function may provide a more accurate fit \citep{eya2024pulsar}. Although the event rate of glitches decrease with pulsar ages, there is not good linear relationship. In our sample, the event rate of clusters indeed has a linear decrease of pulsar ages. This result confirms the fact that the older pulsars have less glitch to happen. 
	\item \citet{melatos2008avalanche} proposed that most pulsar glitches follow a power-law distribution, with waiting times characterized by an exponential distribution. \citet{melatos2018size} and \citet{carlin2019generating} provided a more detailed and quantitative description of this phenomenon based on state-dependent Poisson processes. \citet{fuentes2019glitch} found that the best fit for the size distribution of pulsars PSR J0205$ + $6449, J0631$ + $1036, and B1737$ - $30 is a power-law distribution.  \citet{gao2024comparative} also suggested that PSR B1737$ - $30 exhibits scale invariance within a $q$-Gaussian distribution. These properties were ususally thought to like earthquakes behaviour and can also be extended to other astrophysical phenomena, such as solar flares and repeating fast radio bursts \citep{peng2023scale,totani2023fast}. However, given the observed variations in distributions among different pulsars, adopting the perspective of glitch clusters instead of individual glitches allows for a unified Gaussian fitting of both size and interval time distributions. At the cluster scale, this approach maintains quasi-periodic scale invariance.
\end{enumerate}

The exact causes of glitches and the resulting processes, such as their origin and triggering mechanism are unsolved problems. Whether glitches recur quasi-periodically or not may be a key issue for the understandings of a glitch physics. There appear to be two class of glitches. However, it become possible to bridge the gap between the two classes of glitches, when we here analyze the periodic representation of glitches on longer timescales. The period of glitch cluster may represent the scale that connects the so-called random glitches to a periodic ones. This leads to a possible unified description of diverse glitches. The identified timescale of the glitch cluster periods provides new insights into the relationship between the coupling constant $ G $ and the superfluid reservoir.

Since not all the accumulated pressure within the pulsar is fully released through a glitch, the effective capacity of the angular momentum reservoir involved in glitches appears to be greater than what can be calculated from the observed activity. While the specific relationship between waiting time and glitch size remains uncertain, we argue that longer waiting times result in a broader accumulation and potential release of pressure within pulsars. Consequently, the occurring likelihood of large glitches increases with longer waiting times.

We have adopted a relatively simple and practical grouping method because of not very high number of events. For pulsars with more extensive datasets, future research could involve more refined and reasonable grouping methods to achieve potentially better results. Additionally, considering the entire lifecycle of pulsars, it is worth investigating whether possible physical mechanisms could lead to transformations in the glitch characteristics among different pulsars.

\begin{acknowledgments}
We thank Dr W-H Wang for his useful discussions. This work is supported by the National SKA Program of China under grant No.2020SKA0120300 and National Nature Science Foundation of China under grant No.12033001.
\end{acknowledgments}

\bibliography{References}{}
\bibliographystyle{aasjournal}

\end{document}